\documentclass[twocolumn]{aastex631}
\usepackage{booktabs}
\usepackage{verbatimbox}

\shorttitle{Using the antenna impedance to estimate soil parameters for MIST}
\shortauthors{Altamirano et al.}

\begin{document}

\title{Using the Antenna Impedance to Estimate Soil Electrical Parameters for the MIST Global 21 cm Experiment}

\author[0009-0004-4278-5261]{Cinthia Altamirano}
\affiliation{Departamento de Ingenier\'ia El\'ectrica, Universidad Cat\'olica de la Sant\'isima Concepci\'on, Alonso de Ribera 2850, Concepci\'on, Chile; \href{mailto:caltamirano@ucsc.cl}{caltamirano@ucsc.cl}}


\author[0000-0001-8468-9391]{Ricardo Bustos}
\affiliation{Departamento de Ingenier\'ia El\'ectrica, Universidad Cat\'olica de la Sant\'isima Concepci\'on, Alonso de Ribera 2850, Concepci\'on, Chile}

\author[0000-0002-3287-2327]{Raul A. Monsalve}
\affiliation{Space Sciences Laboratory, University of California, Berkeley, CA 94720, USA}

\affiliation{School of Earth and Space Exploration, Arizona State University, Tempe, AZ 85287, USA}

\affiliation{Departamento de Ingenier\'ia El\'ectrica, Universidad Cat\'olica de la Sant\'isima Concepci\'on, Alonso de Ribera 2850, Concepci\'on, Chile}

\author[0000-0002-8782-2175]{Silvia E. Restrepo}
\affiliation{Departamento de Ingenier\'ia El\'ectrica, Universidad Cat\'olica de la Sant\'isima Concepci\'on, Alonso de Ribera 2850, Concepci\'on, Chile}

\affiliation{Centro de Energ\'ia, Universidad Cat\'olica de la Sant\'isima Concepci\'on, Alonso de Ribera 2850, Concepci\'on, Chile}

\author[0009-0008-9653-6104]{Vadym Bidula}
\affiliation{Department of Physics and Trottier Space Institute, McGill University, Montr\'eal, QC H3A 2T8, Canada}

\author[0000-0002-7971-3390]{Christian H. Bye}
\affiliation{Department of Astronomy, University of California, Berkeley, CA 94720, USA}

\author[0000-0002-4098-9533]{H. Cynthia Chiang}
\affiliation{Department of Physics and Trottier Space Institute, McGill University, Montr\'eal, QC H3A 2T8, Canada}

\affiliation{School of Mathematics, Statistics, \& Computer Science, University of KwaZulu-Natal, Durban, South Africa}

\author[0009-0002-9727-8326]{Xinze Guo}
\affiliation{Space Sciences Laboratory, University of California, Berkeley, CA 94720, USA}

\author[0009-0003-3736-2080]{Ian Hendricksen}
\affiliation{Department of Physics and Trottier Space Institute, McGill University, Montr\'eal, QC H3A 2T8, Canada}

\author[0009-0005-1658-6071]{Francis McGee}
\affiliation{Department of Physics and Trottier Space Institute, McGill University, Montr\'eal, QC H3A 2T8, Canada}

\author[0000-0001-8616-0854]{F. Patricio Mena}
\affiliation{National Radio Astronomy Observatory, Charlottesville, VA 22903, USA}

\author[0009-0001-5711-1153]{Lisa Nasu-Yu}
\affiliation{Department of Physics and Trottier Space Institute, McGill University, Montr\'eal, QC H3A 2T8, Canada}

\author[0000-0001-6903-5074]{Jonathan L. Sievers}
\affiliation{Department of Physics and Trottier Space Institute, McGill University, Montr\'eal, QC H3A 2T8, Canada}

\affiliation{School of Chemistry and Physics, University of KwaZulu-Natal, Durban, South Africa}

\author[0000-0003-1602-7868]{Nithyanandan Thyagarajan}
\affiliation{Commonwealth Scientific and Industrial Research Organisation (CSIRO), Space \& Astronomy, P. O. Box 1130, Bentley, WA 6102, Australia}

\received{2025 April 2}
\revised{2025 July 8}
\accepted{2025 August 11}
\published{2025 August 29}




\begin{abstract}
Radio experiments trying to detect the global $21$~cm signal from the early Universe are very sensitive to the electrical properties of their environment. For ground-based experiments with the antenna above the soil it is critical to characterize the effect from the soil on the sky observations. This characterization requires estimating the soil's electrical conductivity and relative permittivity in the same frequency range as the observations. Here we present our initial effort to estimate the conductivity and relative permittivity of the soil using the impedance of an antenna mounted at a distance above the surface. In this technique, the antenna used for soil characterization is the same as the antenna used for sky observations. To demonstrate the technique we use the antenna of the MIST global $21$~cm experiment. We measured the antenna impedance at three sites in the Greater Concepci\'on area, Chile. The measurements were done between $25$ and $125$~MHz, matching the range used by MIST for sky observations. The soil parameters were estimated by fitting the impedance measurements with electromagnetic simulations of the antenna and soil. In this initial effort the soil was modeled as homogeneous. The conductivity at the three sites was found to be between $0.007$ and $0.049$~Sm$^{-1}$, and the relative permittivity between $1.6$ and $12.7$. The percent precision of the estimates at $68\%$ probability is, with one exception, better (lower) than $33\%$. The best-fit simulations have a better than $10\%$ agreement with the measurements relative to the peak values of the resistance and reactance across our frequency range. For MIST, these results represent a successful proof of concept of the use of the antenna impedance for soil characterization, and are expected to significantly improve in future implementations.
\end{abstract}




\section{Introduction} 

Ground-based experiments trying to detect the sky-averaged, or global, $21$~cm signal from the early Universe are sensitive to the electrical properties of the soil at the observation site \citep{bradley2019,mahesh2021,spinelli2022,monsalve2024a,monsalve2024b,pattison2025}. The global $21$~cm signal, produced during the Universe's dark ages, cosmic dawn, and epoch of reionization, should be detectable at frequencies $\sim$$5$--$200$~MHz \citep{pritchard2012}. At these frequencies, the system temperature of radio receivers is dominated by the contribution from astrophysical foregrounds, which is at least four orders of magnitude larger than the $21$~cm signal \citep{helmboldt2009,guzman2011,panchenko2013,degasperin2020,sasikumarraja2022}. Due to this high system temperature, detecting the global $21$~cm signal requires the calibration of the experiment's instrumental response, including effects from the soil, with an accuracy of at least one part in $10^4$. 

\begin{figure*}
\centering
\includegraphics[width=\hsize]{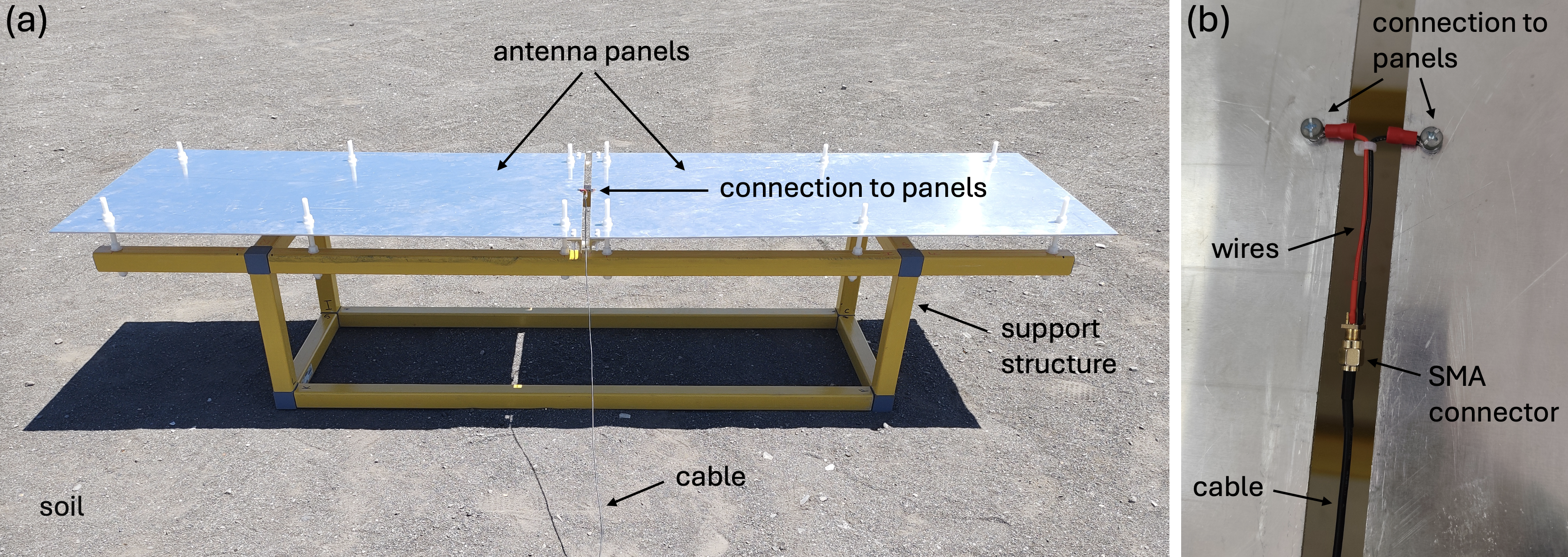}
\caption{MIST blade dipole antenna used to demonstrate the soil characterization technique. The antenna consists of two aluminum panels of size $1.2$~m~$\times$~$60$~cm~$\times$~$3$~mm, with a horizontal separation of $2$~cm. The panels are supported at a height of $50$~cm by a structure made out of fiberglass and plastic. The impedance of the antenna is measured in the gap between the two panels using an adapter. The adapter consists, at one end, of two copper wires bolted to each of the antenna panels and, at the other end, a female SMA connector. A $20$~m coaxial cable with SMA connectors is used to measure the antenna impedance with a VNA.}
\label{figure_antenna}
\end{figure*}

Several ground-based global $21$~cm experiments, such as EDGES \citep{bowman2018}, PRI$^Z$M \citep{philip2019}, LEDA \citep{spinelli2021}, REACH \citep{deleraacedo2022}, SARAS \citep{singh2022}, MIST \citep{monsalve2024a}, and RHINO \citep{bull2024}, have been designed around a single, wide-beam antenna. When this type of antenna operates close to the soil relative to the observation wavelength, the electromagnetic parameters, including the impedance and beam, are directly impacted by the soil's electrical properties \citep{smithrose1933, smithrose1935}. To reduce the effect of the soil on the antenna performance, EDGES, PRI$^Z$M, LEDA, and REACH, have opted for operating their antennas above a metal ground plane. In the case of EDGES, PRI$^Z$M, and LEDA, the ground plane is resting on the soil, while for REACH it is lifted above the soil \citep{deleraacedo2022}. 

Although finite metal ground planes significantly reduce the sensitivity of the antenna to the soil, they can also introduce frequency dependence, or chromaticity, into the antenna beam, which is typically unwanted. This chromaticity is produced by signal reflections from the edges of the ground plane, which represent an electrical discontinuity between the metal and the soil \citep{rogers2022}. For the ground plane sizes used by current instruments, of order tens of meters on a side, this beam chromaticity generates spectral structure in the measured antenna temperature on scales comparable to global $21$~cm signal models \citep{pritchard2012, mahesh2021, spinelli2022}. If uncorrected or imperfectly accounted for, this structure could be confused for the cosmological signal or complicate its extraction. Furthermore, problems can occur due to imperfections in the ground plane such as unintended slots in the metal, which would generate resonances \citep{bradley2019}, and bumps, which would produce additional reflections \citep{rogers2022}. These imperfections can be difficult to diagnose and their effects difficult to estimate and remove during data analysis. Finally, since finite metal ground planes do not eliminate the sensitivity to the soil, it is necessary to characterize the soil and account for its impact even if a ground plane is used \citep{monsalve2017, bowman2018, mahesh2021, monsalve2021, deleraacedo2022, spinelli2022, pattison2025}.

To avoid the negative effects mentioned above, MIST has taken the approach of observing the sky with the antenna above the soil but without using a metal ground plane \citep{monsalve2024a}. This approach has the trade-off of folding the full effect from the soil into MIST’s electromagnetic performance. Although the magnitude of the soil effect is larger in an absolute sense, it is possible that the spectral scales affected in this approach represent a more favorable scenario than when using a metal ground plane. Whether or not this approach yields better results will depend on the specific characteristics of the soil and how we account for its effects during analysis. In our implementation, the effect from the soil on the antenna impedance is directly determined from the impedance measurements conducted in the field \citep{monsalve2024a}. On the other hand, the effect on the beam directivity and antenna losses is estimated using electromagnetic simulations of the instrument that include the soil. For these simulations, it is critical to have accurate knowledge of the soil's electrical characteristics.

In \citet{singh2022} SARAS measured the sky from a lake, thus avoiding the use of a large metal ground plane similarly to MIST. Due to its homogeneity, water is an ideal medium to use as a ``soil'' in real observations as well as in the context of simulations. However, operating over a sufficiently large body of water poses serious logistical challenges. These challenges are increased if the observation site is in a remote radio-quiet location. In contrast, operating above soil typically provides the safe and stable conditions required for the careful study of systematic effects and for achieving a high signal-to-noise ratio in the sky observations.

To maximize the accuracy of the electromagnetic simulations, MIST's observation sites are chosen optimizing for surface flatness and homogeneous composition. Although inhomogeneities are unavoidable in natural soils, their effect might become subdominant when averaged over hundreds of meters around the antenna. In this case, simulating the soil as homogeneous or stratified could represent a sufficiently accurate approximation. This possibility is one of MIST's research hypotheses. Leveraging MIST's small size and portability, our experimental program includes operating from different sites to improve the understanding and mitigation of soil effects \citep{monsalve2024a}.

In this paper, we demonstrate a technique for the electrical characterization of the soil at sites used for observations of the global $21$~cm signal. In this technique, the electrical properties of the soil are inferred from the impedance of the antenna used for sky observations. The parameters determined by this technique are the soil's bulk electrical conductivity and relative permittivity. These parameters are estimated by fitting impedance measurements with impedance models obtained from electromagnetic simulations of the antenna and soil. 

We demonstrate the technique using the antenna from the MIST experiment. Our impedance measurements and simulations span the frequency range $25$--$125$~MHz, matching the range used by MIST for observations of the sky spectrum \citep{monsalve2024a}. Measurements of the antenna impedance are an essential part of MIST's calibration program. They are conducted to calibrate the impedance mismatch between the antenna and the receiver input. However, here we explore the electrical characterization of the soil, also fundamental for MIST, as an additional use of the antenna impedance. Characterizing the soil using the antenna impedance could eliminate the need for additional measurements and equipment considered for this task.

As indicated in \citet{ieee2011}, using the impedance of an antenna located above the surface is a valid non-invasive approach for the electrical characterization of the soil at radio frequencies. This technique has been used for decades to characterize the electrical properties of materials in an antenna's vicinity. Early results were obtained by comparing the measured impedance with analytical models \citep[e.g.,][]{bhattacharyya1963,wait1969,smith1974,wong1977,nicol1988,he1992}. In several recent efforts the measurements have been compared with models from electromagnetic simulations \citep[e.g.,][]{wakita2000,lenlereriksen2004,berthelier2005,legall2006}. Our implementation of this technique represents an improvement on the latter works thanks to the availability of more powerful computers.

\section{Antenna and Measurements}

\subsection{Blade Dipole Antenna}
\label{section_antenna}
Panel~(a) of Figure~\ref{figure_antenna} shows the blade dipole antenna used in this work. This antenna is a replica of the antennas used by MIST for the sky measurements described in \citet{monsalve2024a}. The antenna consists of two solid aluminum panels of $1.2$~m in length and $60$~cm in width, mounted horizontally $50$~cm above the soil. The panel thickness is $3$~mm and the horizontal separation between the panels is $2$~cm. The antenna is supported by a structure made out of fiberglass tubes and angles, as well as plastic connectors, screws, and nuts.

The antenna is fed through an adapter located in the gap between the panels. This adapter is shown in panel~(b) of Figure~\ref{figure_antenna}. The adapter consists, at one end, of two copper wires bolted to each of the antenna panels and, at the other end, a female subminiature version A (SMA) connector. The antenna impedance is measured using a coaxial cable connected to the adapter. The adapter does not provide direct current (DC) isolation between the panels and the cable. Specifically, one panel has DC continuity with the center conductor of the cable and the other panel has DC continuity with the outer conductor. As it leaves the antenna, the cable runs in the middle of the gap between the panels and perpendicular to the dipole excitation axis.

The full MIST instrument also includes a receiver box installed under the antenna, which contains the electronics and batteries, and a balun connected between the antenna and the receiver box \citep{monsalve2024a}. In this initial study, however, we test the soil characterization technique using the impedance of the antenna by itself, without the receiver box and balun.

\begin{figure}
\centering
\includegraphics[width=\hsize]{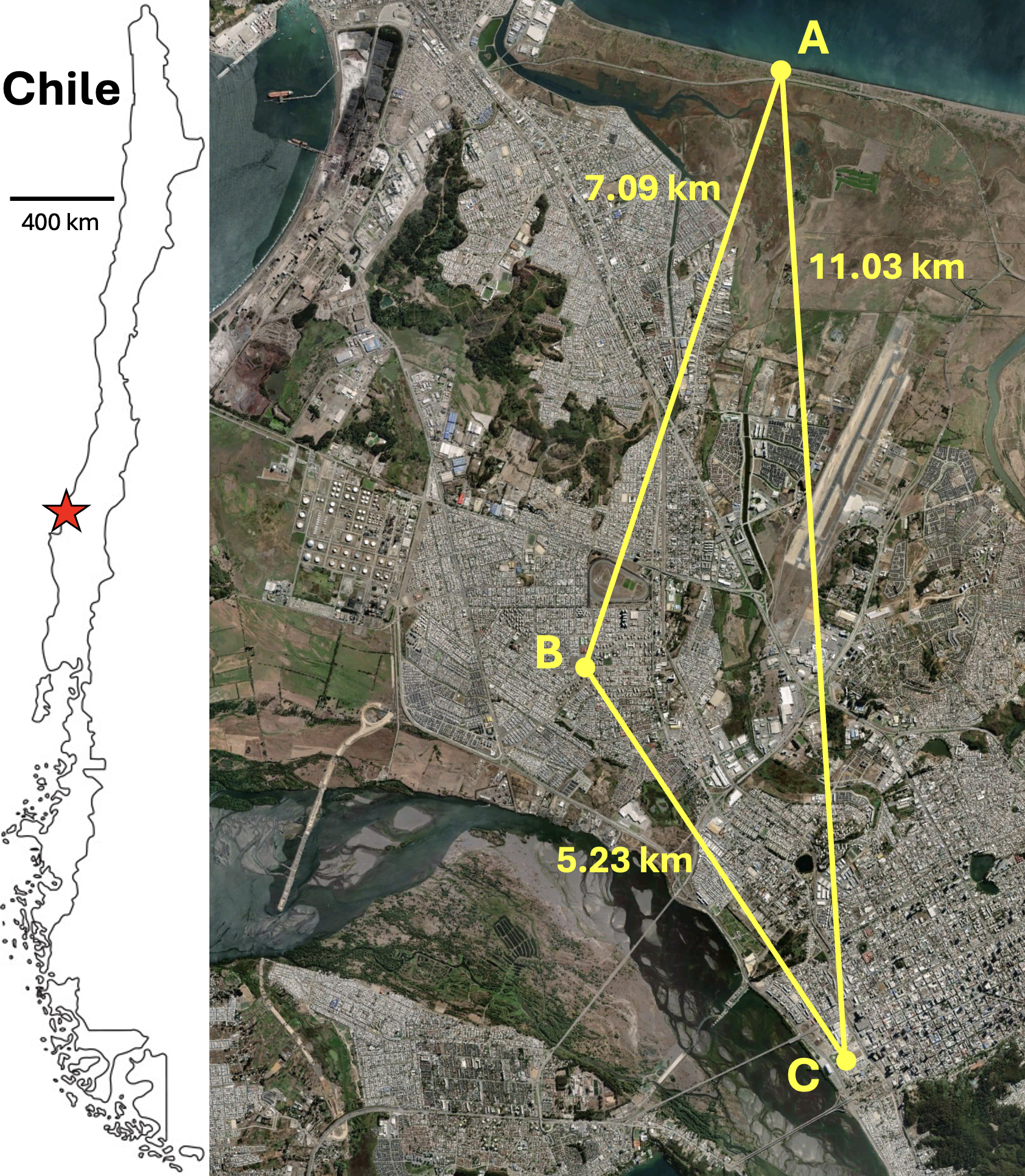}
\caption{Location of the test sites in the Greater Concepci\'on area, Chile.}
\label{figure_sites}
\end{figure}

\begin{table}
\caption{Test sites in the Greater Concepci\'on area, Chile.}
\label{table_sites}
\centering
\begin{tabular}{cll} 
\hline
\hline
\noalign{\smallskip}
Site &	Coordinates & Surface composition 	\\
\noalign{\smallskip}
\hline
\noalign{\smallskip}
A &	$36^{\circ}43'56.3''$~S, $73^{\circ}04'14.9''$~W & sand, grass			\\
B &	$36^{\circ}47'34.2''$~S, $73^{\circ}05'45.2''$~W & sand, gravel, no grass	\\
C &	$36^{\circ}49'53.5''$~S, $73^{\circ}03'45.2''$~W & clay, grass   \\
\noalign{\smallskip}
\hline
\end{tabular}
\end{table}

\subsection{Test Sites}
Measurements were conducted at three sites in the Greater Concepci\'on area, Chile, on 2023 November~14. The sites are shown in Figure~\ref{figure_sites} and their coordinates are listed in Table~\ref{table_sites}. Sites~A, B, and C, are located on a beach, a football field, and a park, respectively. These sites were selected because they have similar characteristics to the sites used by MIST for sky observations \citep{monsalve2024a}. At the three sites the surface is level to within a few degrees and relatively flat, with vertical variations within $\sim$$5$~cm ($25$~cm) inside a $\sim$$2$~m ($20$~m) radius from the antenna. Within a $\sim$$20$~m radius, there are no tall plants, trees, or buildings that could cause signal reflections. At Site~A, the antenna is also more than $20$~m away from the sea water. Table~\ref{table_sites} gives a general description of the soil composition at the surface.

\subsection{Measurement Setup}
The measurements were done with a NanoVNA-H\footnote{\url{https://nanovna.com/}} vector network analyzer (VNA), in the frequency range $25$--$125$~MHz, with a $1$~MHz resolution. The data provided by the VNA correspond to the complex reflection coefficient, $\Gamma$, which we convert to impedance, $Z$, using the following equation \citep{pozar2005}:

\begin{equation}
Z = Z_0 \frac{1 + \Gamma}{1 - \Gamma},
\label{equation_impedance}
\end{equation}

\noindent where $Z_0=50$~$\Omega$ is the characteristic impedance of the measurement setup.

We connected the VNA to the antenna using a $20$~m RG174 cable with male SMA connectors. The VNA was operated manually and the operator was located by the VNA, $\sim$$20$~m away from the antenna. The measurements were done with the operator sitting on the ground to minimize signal blockages and reflections. Before each antenna measurement, the setup was calibrated using the generic open, short, and load (OSL) SMA standards included with the NanoVNA-H. The calibration was done at the antenna end of the cable and, therefore, removed the unwanted S-parameters of the VNA and cable.

\begin{figure}
\centering
\includegraphics[width=\hsize]{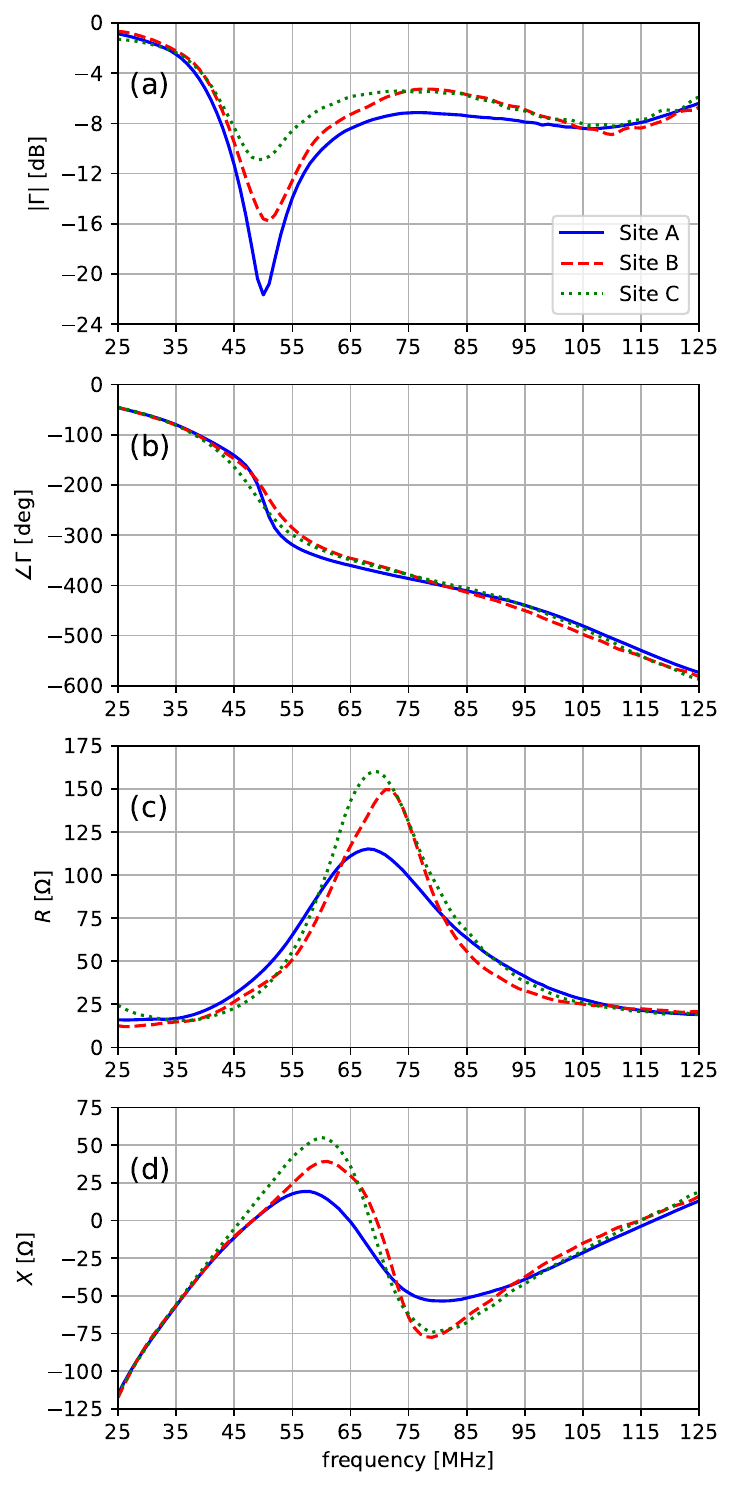}
\caption{Measurements of the MIST antenna at the three test sites. (a) Reflection magnitude. (b) Reflection phase. (c) Resistance. (d) Reactance. The differences in the measurements across the sites are primarily due to differences in the electrical properties of the soils.}
\label{figure_measurements}
\end{figure}

\subsection{Measurements}
\label{section_measurements}
The measurements are shown in Figure~\ref{figure_measurements}. Panels~(a) and (b) show, respectively, the magnitude and phase of the reflection coefficient. Panels~(c) and (d) show, respectively, the resistance ($R$, real part of the impedance) and reactance ($X$, imaginary part of the impedance). The differences observed between the three sites are primarily due to differences in the electrical properties of the soils. The common features across sites include: (1) a dip in $|\Gamma|$ at $\sim$$50$~MHz; (2) a fast transition in $\angle \Gamma$ at $\sim$$50$~MHz; (3) a peak in $R$ at $\sim$$65$--$75$~MHz; and (4) a fast transition in $X$ at $\sim$$60$--$80$~MHz. Similar features are observed in the measurements done by MIST in California, Nevada, and the Canadian High Arctic \citep{monsalve2024a}.

The measurement noise in the resistance and reactance has a sample standard deviation $< 0.2$~$\Omega$, which was estimated from residuals resulting from polynomial fits to the data. For this initial study we did not estimate the systematic uncertainty in the measurements, which is primarily associated with the accuracy and stability of the OSL calibration. We expect this systematic uncertainty to dominate the measurement uncertainty budget. The determination of this uncertainty will be one of our priorities in future implementations of this technique.

\begin{figure}
\centering
\includegraphics[width=\hsize]{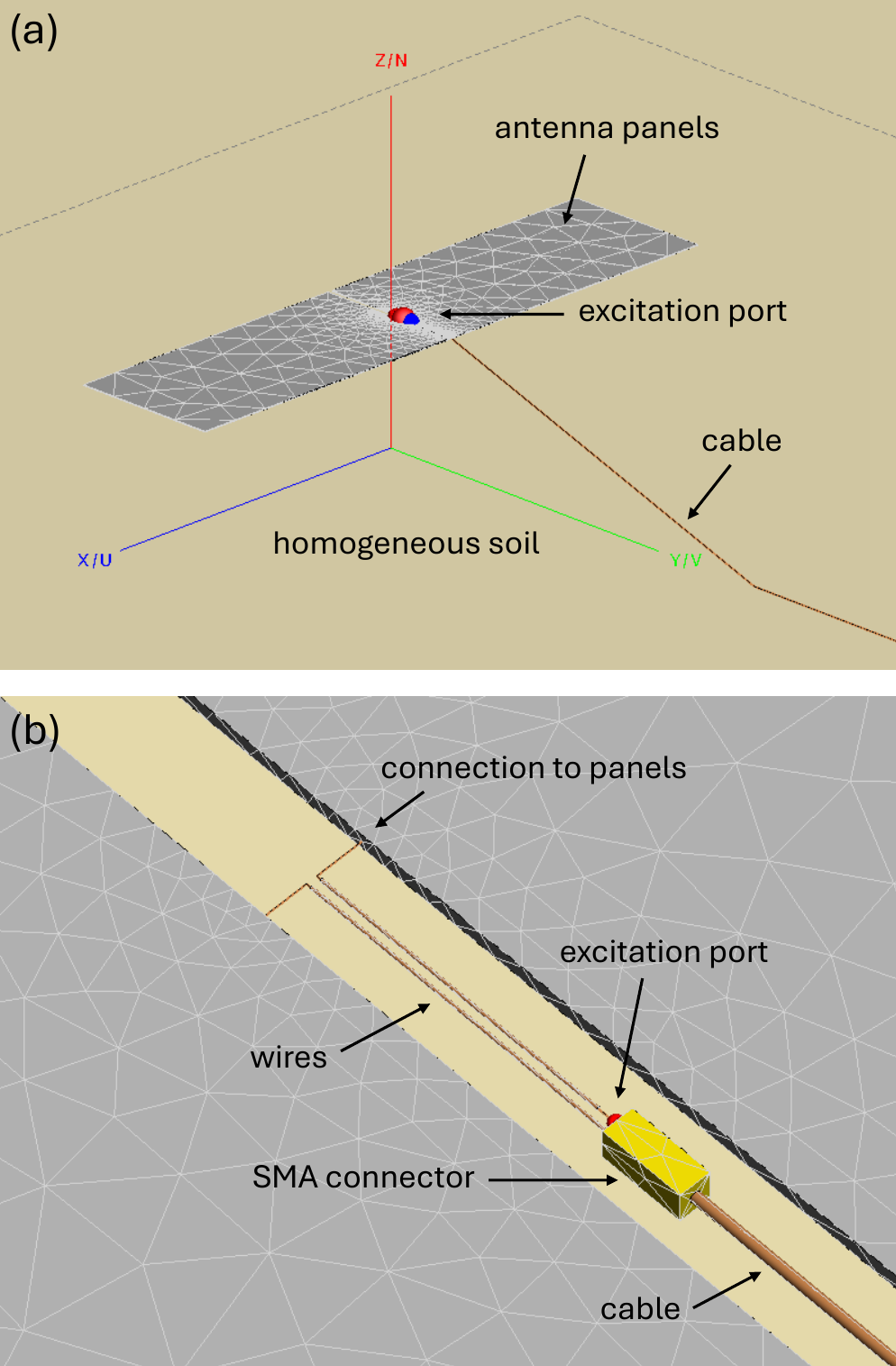}
\caption{Geometry of the Feko simulations. The geometry includes the soil, $20$~m cable, antenna panels, and wires and SMA connector in the gap between the panels. The soil was simulated as homogeneous, and infinite in the horizontal directions and in depth. The meshing seen in panel~(b) on the geometry (except on the soil) corresponds to a discretization of the surfaces for the computation of currents and fields with the method of moments.}
\label{figure_feko}
\end{figure}

\section{Electromagnetic Simulations}
The electromagnetic simulations of the antenna were computed with the Feko software\footnote{\url{www.altair.com/feko}} and its implementation of the method of moments. The geometry of the simulations is shown in Figure~\ref{figure_feko} and described in detail in Appendix~\ref{section_appendix_feko}. The geometry includes the soil, the antenna panels, the wires and SMA connector that form the adapter in the gap between the panels, and the $20$~m cable starting in the gap and continuing toward the VNA. The geometry was designed using nominal dimensions of the antenna panels and $20$~m cable, and laboratory measurements of the adapter.

During a preliminary exploration, the geometry only consisted of the antenna panels, an idealized excitation port, and the soil. With this simple geometry, however, no soil parameter combinations could produce a reasonable agreement with the measurements (Appendix~\ref{section_appendix_simple_geometry}). For this reason we incorporated the additional features (Appendix~\ref{section_appendix_full_geometry}), after which the agreement between the simulations and the data improved significantly.

The assembly of the antenna, adapter, and cable was replicated at the three test sites as closely as possible, and in this paper we use the same geometry for the simulations corresponding to the three sites. The simulations were conducted in the range $25$--$125$~MHz using double precision calculations and \verb~fine~ automatic mesh size.

\subsection{Soil Model}
The soil was simulated as homogeneous, with a perfectly flat surface, and infinite in the horizontal directions and in depth. The electrical parameters of the soil model are its conductivity ($\sigma$) and relative permittivity ($\epsilon_r$)\footnote{These parameters are components of the complex permittivity of a medium, $\epsilon$, defined as $\epsilon = \epsilon'-j\epsilon''=\epsilon_0\epsilon_r - j\frac{\sigma}{2\pi f}$, where $\epsilon_0$ is the permittivity of vacuum and $f$ is frequency \citep{vonhippel1995, pozar2005}.}. This soil model is a first-order approximation to the real soils at our sites, in which the surface is not perfectly flat and the properties, in general, vary as a function of space. The electrical properties of real soils, in general, also vary as a function of frequency \citep[e.g.,][]{smithrose1933, smithrose1935, hipp1974, hoekstra1974, hallikainen1985, dobson1985, francisca2003, bobrov2015, sutinjo2015} and time \citep{pattison2025}. In this study, $\sigma$ and $\epsilon_r$ were modeled as constant in the range $25$--$125$~MHz. We made this choice under the assumption that any variations of $\sigma$ and $\epsilon_r$ with frequency will have a small effect on the impedance compared to our experimental uncertainties. In Feko, the effect of the soil was incorporated analytically using Sommerfeld integrals as Green’s functions for solving the boundary conditions \citep{davidson2011, mosig2021}.

\begin{figure}
\centering
\includegraphics[width=\hsize]{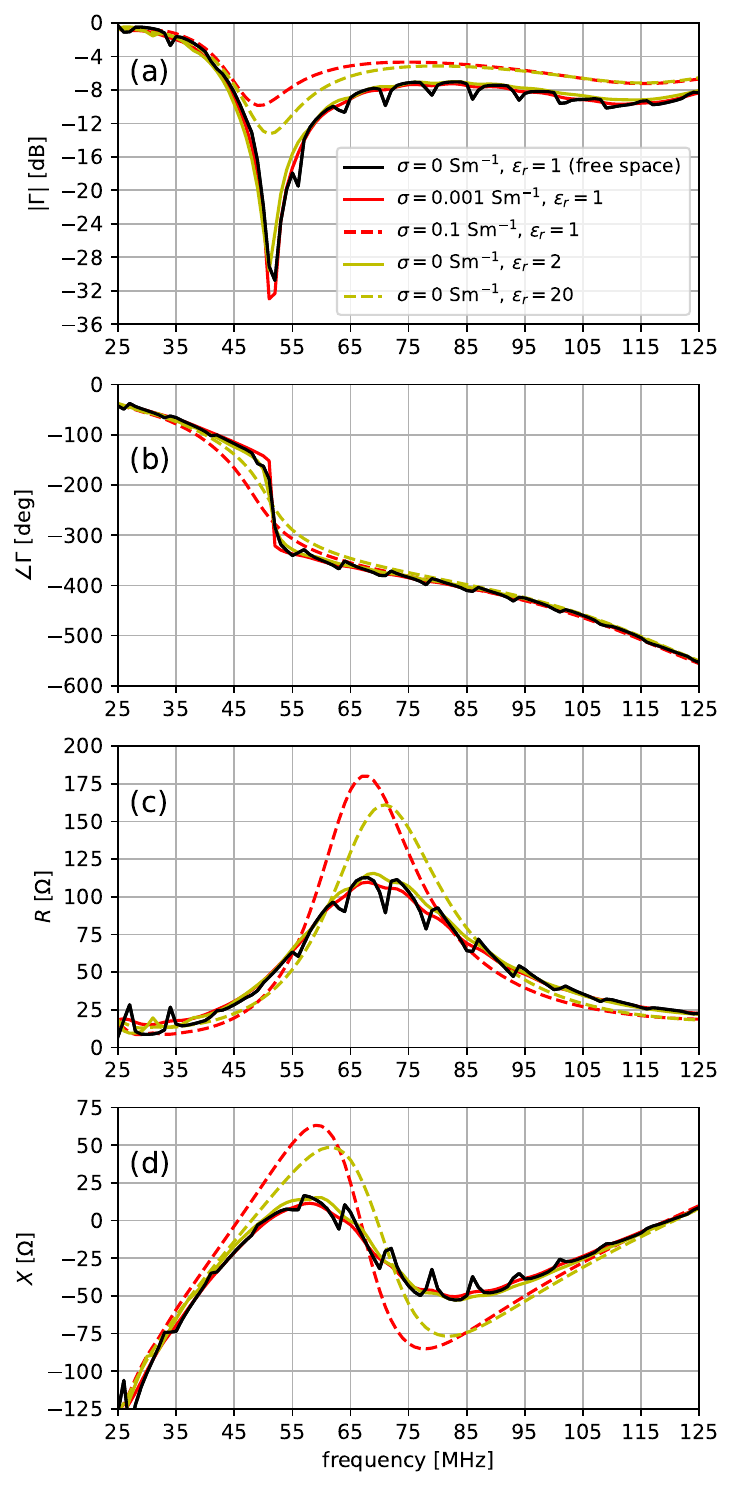}
\caption{Feko simulations conducted to study the sensitivity of $\Gamma$ and $Z$ to the soil parameter values. The reference case (black line) corresponds to the geometry simulated in free space. In the reference case, the structure observed across frequency with a period of $\sim$$7.5$~MHz is produced by reflections from the end of the $20$~m cable. This structure is suppressed when $\sigma$ or $\epsilon_r$ are increased because reflections from the soil, which span a wide range of delays, have a smoothing effect and reduce the impact from the cable alone.}
\label{figure_sensitivity}
\end{figure}

\subsection{Sensitivity to the Soil Parameters}
\label{section_sensitivity}
To develop intuition about the sensitivity of the antenna reflection coefficient and impedance to the soil parameters, in Figure~\ref{figure_sensitivity} we show Feko simulations for five combinations of soil parameter values. The reference parameter combination is $(\sigma,{\epsilon}_r)=(0\;\mathrm{Sm}^{-1},1)$, which is equivalent to free space. Relative to the reference, in two cases we increase $\sigma$ and in the other two we increase $\epsilon_r$. The features across frequency that are common among the simulations are those also identified in the measurements (Section~\ref{section_measurements}); i.e., a dip in $|\Gamma|$, a fast transition in $\angle \Gamma$, a peak in $R$, and a fast transition in $X$. We note that these characteristic features are present even when the antenna is in free space, and can therefore be regarded as intrinsic to the antenna structure instead of imposed by the soil. An additional noteworthy feature observed in the free space simulation is the repetitive structure seen across frequency in $|\Gamma|$, $\angle \Gamma$, $R$, and $X$. This structure has a period of $\sim$$7.5$~MHz, which in free space corresponds to a wavelength or travel distance of $\sim$$40$~m. The structure is therefore attributed to reflections from the end of the $20$~m cable included in the full geometry. When $\sigma$ or $\epsilon_r$ increase compared to free space, the periodic structure is suppressed. This behavior occurs because the soil surface also produces reflections, which span a wide range of delays and have a smoothing effect across frequency. Figure~\ref{figure_sensitivity} also shows that increasing $\sigma$ or $\epsilon_r$ produces a higher $|\Gamma|$, a smoother transition in $\angle \Gamma$, a narrower and higher peak in $R$, and a larger amplitude in the transition of $X$. Finally, we note the following patterns in the impedance. Increasing $\sigma$ ($\epsilon_r$) shifts the peak of $R$ to lower (higher) frequencies. Regarding $X$, increasing $\sigma$ shifts the high-frequency knee of the transition (dip at $\sim$$80$~MHz) to lower frequencies. Conversely, increasing $\epsilon_r$ shifts the low-frequency knee of the transition (peak at $\sim$60~MHz) to higher frequencies.

\section{Parameter Estimation}

We estimate the soil conductivity and relative permittivity at our test sites by fitting the measured antenna impedances with models from Feko simulations. As described in Section~\ref{section_measurements}, in this initial implementation of the technique we do not know the dominant measurement uncertainties. For this reason, instead of minimizing chi squared, in the fits we minimize the sum of squared deviations (SSD), 

\begin{equation}
\mathrm{SSD} = \sum_{i=1}^N{(d_i-m_i)^2},
\end{equation}

\noindent where $d$ represents the data, $m$ is the impedance model from a Feko simulation, and $i$ is the index of the $N$ points available across the frequency range. The best-fit soil parameters, which we call $\hat{\sigma}$ and $\hat{\epsilon}_r$, are those that produce the lowest SSD, denoted by SSD$^{\star}$.

Carrying out Feko simulations in the range $25$--$125$~MHz at the $1$~MHz resolution of the measurements leads to computation times of $\sim$$40$~minutes per simulation on our 64-core AMD Ryzen Threadripper PRO 5995WX computer. These computation times are impractical in a pipeline requiring a rigorous sampling of the soil parameter space. The computation time roughly scales with number of frequency points requested. To decrease the computation time, we reduced the resolution of the simulations to $10$~MHz and, thus, the number of points requested from $101$ to $11$. Using $11$ complex data points evenly spaced across $25$--$125$~MHz effectively captures the main features of the antenna impedance and is sufficient for our analysis. With this configuration, a simulation for a single combination of soil parameters is computed in $3.5$~minutes. At each of the $11$ frequency channels where data and model are available, the resistance and reactance are treated as two independent points and, therefore, $N=22$.

Since we do not know the measurement uncertainties when conducting the fits, we cannot compute the data likelihood and posterior probability density function (PDF) of $\sigma$ and $\epsilon_r$ with high fidelity. Therefore, it is not justified to fit the soil parameters using techniques such as nested sampling or MCMC, which are designed to map the PDFs and would likely require thousands of computationally expensive Feko simulations. We instead use a method that conducts a global search to determine the best-fit values and requires only a few hundred simulations. This method is described in Appendix~\ref{section_appendix_fitting_strategy}. At the core of the method is the simplicial homology global optimization (SHGO) algorithm. In this method, we initially fit the soil parameters considering a wide parameter space: $0$--$1$~Sm$^{-1}$ for $\sigma$ and $1$--$20$ for $\epsilon_r$. This space includes the values from all the geophysical materials that could realistically be present in the soil at our three test sites \citep{reynolds2011}. This initial step identifies the general regions where the best-fits lie but does not provide accurate estimates. To increase the fitting accuracy, we carry out a second fitting step where the global search is conducted within a smaller parameter space. This smaller space is chosen based on the results from the first step. SHGO does not directly provide uncertainties in the best-fit parameters. However, we determine first-order parameter uncertainties in a separate procedure using the samples acquired during the fits. This procedure is described in Section.~\ref{section_best_fit_uncertainty}.

\section{Results}

\subsection{Best-fit soil parameters and impedance simulations}
\label{section_best_fits}

For Sites~A, B, and C, the best-fit parameter values found with SHGO are $(\hat{\sigma},\hat{\epsilon}_r)=(0.0070\;\mathrm{Sm}^{-1},1.63)$, $(0.0165\;\mathrm{Sm}^{-1},9.77)$, and $(0.0494\;\mathrm{Sm}^{-1},12.73)$, respectively. The corresponding $\mathrm{SSD}^{\star}$ are $513.3$, $1815.5$, and $825.9$~$\Omega^2$. The total number of Feko simulations conducted to determine the best-fit values at each of the three test sites was $\leq 602$. More details of the fits are provided in Appendix~\ref{section_appendix_fitting_strategy}. The best-fit parameter values are shown in Table~\ref{table_best_fits} including uncertainty ranges, which are computed as described in Sec.~\ref{section_best_fit_uncertainty}.

Figure~\ref{figure_residuals} compares the measured impedances with the best-fit simulations. The data and simulations are shown at the common $10$~MHz resolution. The simulations show good general agreement with the data, with residuals that are within $\pm 25$~$\Omega$, and for most of the points within $\pm 10$~$\Omega$, especially for Sites~A and C. We also quantify the residuals using the root mean square deviation (RMSD) between the best-fit models and the data; specifically, $\mathrm{RMSD}^{\star}=\sqrt{\mathrm{SSD}^{\star}/N}$. The RMSD$^{\star}$ considers both the resistance and reactance, and is $4.8$, $9.1$, and $6.1$~$\Omega$ for Sites A, B, and C, respectively. These values represent a better than $10\%$ agreement between best-fit simulations and data when using as a reference the peak absolute values of the resistance and reactance across our frequency range, which are $>100$~$\Omega$ for the three sites. When considering the origin of the residuals, we cannot rule out the possibility that at least part of them is due to the inadequacy of our homogeneous soil models to simulate the real soils. Unfortunately, our interpretations are fundamentally restricted by the lack of information about our measurement uncertainties. The quantification of these uncertainties should therefore be prioritized in future implementations of this technique.

\begin{table}
\caption{Best-fit values, $68\%$ uncertainties, and percent precision for the soil conductivity and relative permittivity at the three test sites.}
\label{table_best_fits}
\centering
\begin{tabular}{ccc}
\hline
\hline
\noalign{\smallskip}
Site & $\hat{\sigma}\pm\delta_{\sigma}$~[Sm$^{-1}$] & $\hat{\epsilon}_r\pm\delta_{\epsilon_r}$ \\
\noalign{\smallskip}
\hline
\noalign{\smallskip}
A       & $0.0070^a \pm 0.0014^b$ ($20\%$)  & $1.6 \pm 0.4$ ($25\%$) \\
B       & $0.017 \pm 0.010$ ($59\%$)  & $9.8 \pm 3.1$ ($32\%$) \\
C       & $0.049 \pm 0.010$ ($20\%$)  & $12.7 \pm 3.2$ ($25\%$)  \\
\noalign{\smallskip}
\hline
\multicolumn{3}{l}{$^a$Best-fit obtained directly with SHGO.}\\
\multicolumn{3}{l}{$^b$$68\%$ uncertainty obtained as described in Section~\ref{section_best_fit_uncertainty}.}
\end{tabular}
\end{table}

\subsection{Uncertainties in the best-fit soil parameters}
\label{section_best_fit_uncertainty}

We estimate first-order uncertainties in the best-fit soil parameters from estimates of the parameters' posterior PDFs. Bayes' theorem indicates that the joint posterior PDF of the parameters given the data is

\begin{equation}
P(\sigma,\epsilon_r|d) \propto \mathcal{L}(d|\sigma,\epsilon_r)\pi(\sigma,\epsilon_r),
\label{equation_bayes}
\end{equation}

\noindent where $\mathcal{L}(d|\sigma,\epsilon_r)$ represents the likelihood of the data given the parameters and $\pi(\sigma,\epsilon_r)$ represents the prior PDF. We compute the posterior PDFs using the samples produced during the SHGO parameter fitting. As a prior we use uniform PDFs that span the parameter spaces selected for the final fitting step with SHGO (Appendix~\ref{section_appendix_fitting_strategy}). Computing $\mathcal{L}$ requires knowledge of the measurement uncertainties. Since we do not have this information we make simplifying assumptions. Specifically, we first assume that the measurements are affected by errors that are random, Gaussian, and uncorrelated across frequency. We further assume that the errors have a frequency-independent standard deviation equal to the RMSD$^{\star}$ reported in Sec.~\ref{section_best_fits}. With these assumptions, the likelihood becomes

\begin{figure*}
\centering
\includegraphics[width=\hsize]{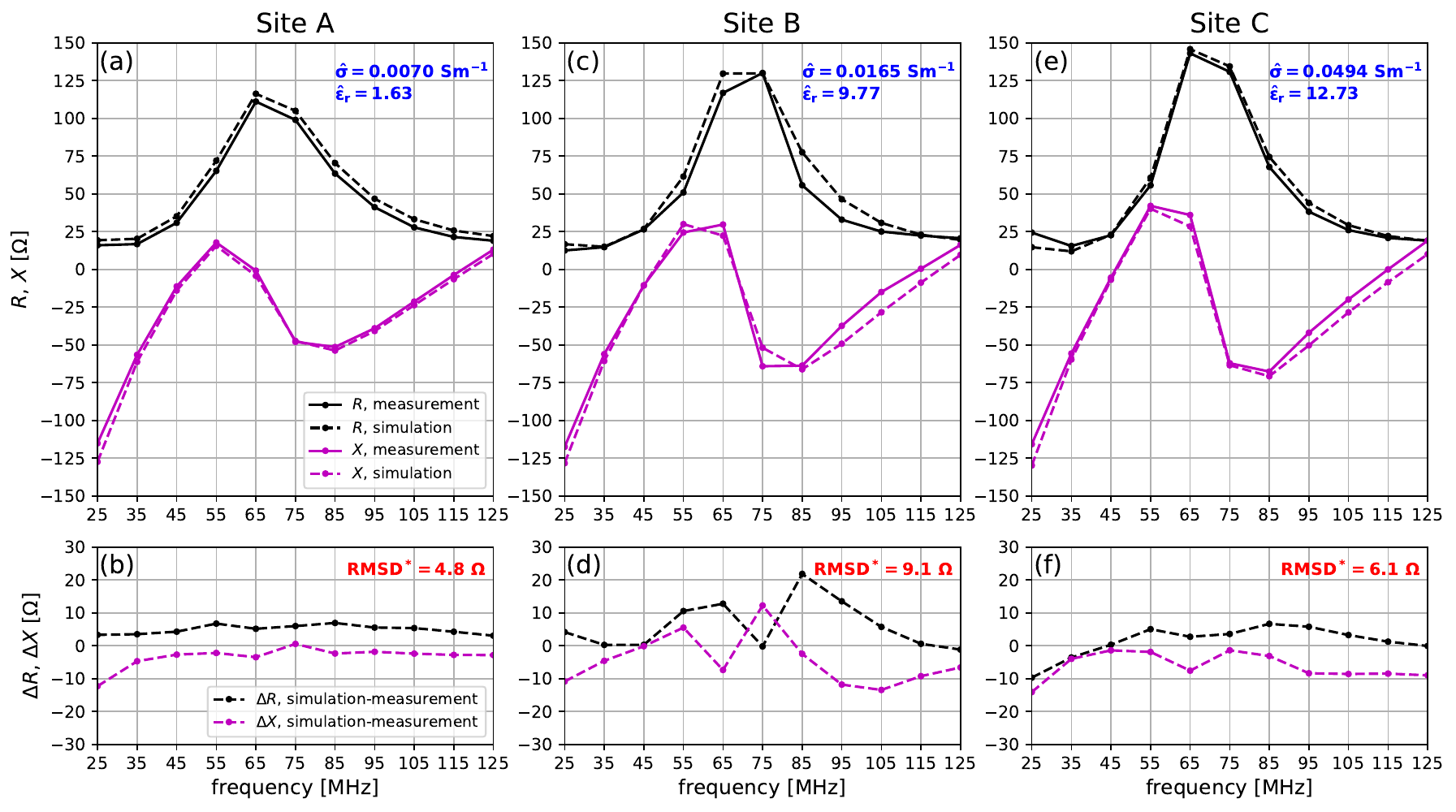}
\caption{Summary of the best-fit results. The top row shows the impedance measurements (solid lines), best-fit simulations (dashed lines), and best-fit soil parameter values (blue font). The bottom row shows the best-fit residuals (dashed lines) and RMSD$^{\star}$ (red font).}
\label{figure_residuals}
\end{figure*}

\begin{equation}
\mathcal{L} \propto \exp\left[-\frac{\mathrm{SSD}}{2(\mathrm{RMSD}^{\star})^2}\right].
\label{equation_likelihood_ssd}
\end{equation}

The structure of the residuals in Figure~\ref{figure_residuals} does not appear random, Gaussian, or uncorrelated across frequency, but we make these assumptions about the errors as an approximation in the absence of independently determined information. We caution, therefore, that the parameter uncertainties determined here should only be regarded as first-order estimates.

\begin{figure*}
\centering
\includegraphics[width=\hsize]{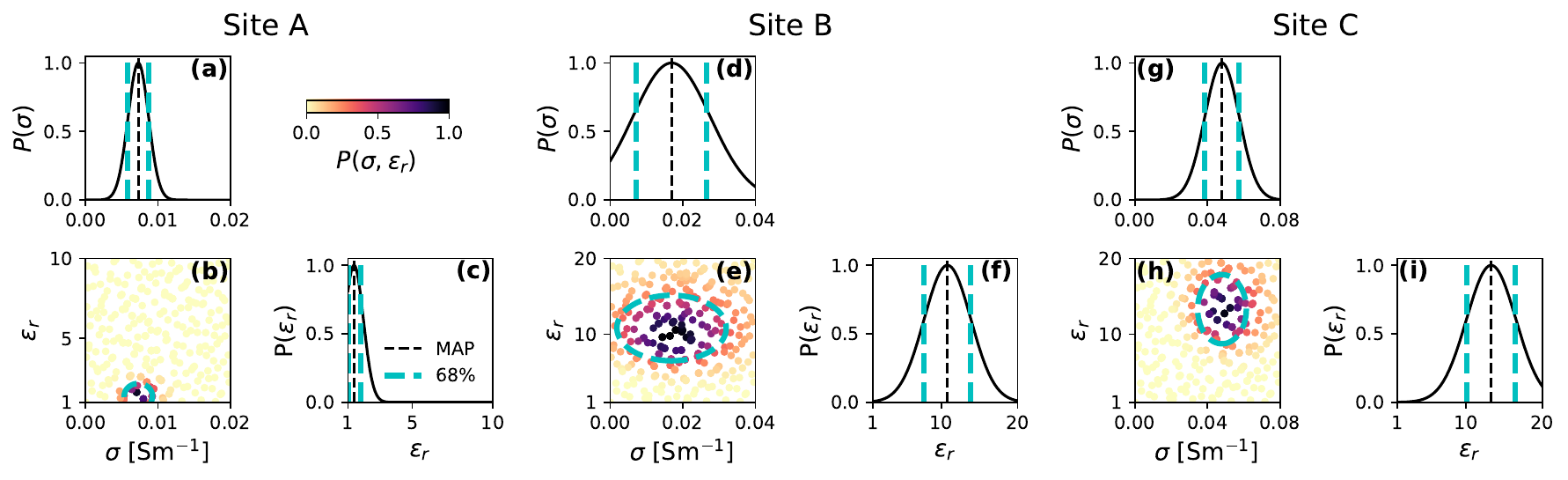}
\caption{Peak-normalized PDFs for $\sigma$ and $\epsilon_r$ obtained assuming that the uncertainty in the impedance measurements is equal to the RMSD$^{\star}$ reported in Sec.~\ref{section_best_fits}. For Site~A, panel~(b) shows the joint PDF for $\sigma$ and $\epsilon_r$, while panels (a) and (c) show the marginalized PDF for each parameter. The $68\%$ probability limits are indicated using the dashed cyan lines. We obtain $68\%$ uncertainty ranges for the best-fit parameters from the $68\%$ limits in the marginalized PDFs. Equivalent information for Sites~B and C is shown in panels (d)-(i). }
\label{figure_pdfs}
\end{figure*}

The posterior PDFs computed for the soil parameters are shown in Figure~\ref{figure_pdfs}. Panels~(b), (e), and (h) show the joint PDFs for Sites~A, B, and C, respectively. The samples shown in the panels come from the SHGO fits. The PDF surfaces revealed by the samples are smooth and contain compact regions of high probability that resemble Gaussians. Joint $68\%$ probability limits were computed after interpolating the joint PDFs at high resolution between the SHGO samples. These limits are shown using the dashed cyan lines. In Figure~\ref{figure_pdfs}, the other panels show marginalized PDFs for each individual parameter. For example, panels~(a) and (c) show, respectively, the marginalized PDFs for $\sigma$ and $\epsilon_r$ at Site~A. The marginalized PDFs were computed from the high-resolution interpolation of the joint PDFs. The $68\%$ uncertainties in the soil parameters, denoted by $\delta_{\sigma}$ and $\delta_{\epsilon_r}$, are determined from the marginalized PDFs; specifically, from the distance between the $68\%$ limits (dashed cyan lines) and the maximum a posteriori (MAP, dashed black lines). 

Our results are shown in Table~\ref{table_best_fits}. For all parameters and sites, the ``+'' and ``-'' $68\%$ ranges from the marginalized PDFs are equal to within $10\%$, and we therefore report as the $68\%$ uncertainty the average of the two. The table also reports the percent precision of the estimates, which for $\hat{\sigma}$ is computed as $100\%\times(\delta_{\sigma}/\hat{\sigma})$. For Site~B, the percent precision of $\hat{\sigma}$ and $\hat{\epsilon}_r$ is, respectively, $59\%$ and $32\%$. For Sites~A and C, the percent precision of $\hat{\sigma}$ and $\hat{\epsilon}_r$ is, respectively, $20\%$ and $25\%$.

\subsection{Comparison with expectations}
In this initial effort to demonstrate the antenna impedance technique we do not attempt to offer a geophysical interpretation of our results shown in Table~\ref{table_best_fits}. We note, however, that our results are in broad agreement with expectations from the soil composition at the test sites. Sand and gravel, found at the surface of Sites~A and B, have reported conductivity $\lesssim0.01$~Sm$^{-1}$ and relative permittivity in the range $\sim$$2$-$30$. For clay, found at the surface of Site~C, the conductivity is in the range $10^{-4}$--$1$~Sm$^{-1}$ and the relative permittivity in the range $\sim$$2$--$40$. Within these ranges, the specific electrical parameter values for soils composed of sand, gravel, or clay are strongly determined by the soil moisture \citep{reynolds2011}. 

To the best of our knowledge, the only independent soil characterization done close to our sites is the one reported in \citet{alegria2013}. They carried out electrical resistivity tomography (ERT) at a site $650$~m east of Site~A. Their site is on the same beach, at coordinates $36^{\circ}44'03.3''$~S, $73^{\circ}03'50.4''$~W. From their measurements they obtained the four-layer conductivity model presented in Table~\ref{table_ERT}.

Our conductivity estimate for Site~A is not expected to match very closely the results from \citet{alegria2013}. First, being $650$~m apart, the two spots do not necessarily have similar soil characteristics. Second, while our result is valid for the range $25$--$125$~MHz, in ERT the resistivity ---inverse of conductivity--- is estimated in the DC regime\footnote{Electrodes are used to inject direct currents into the soil and measure the resulting voltages between different spots \citep{slichter1933,koh1997,ieee2011,herring2023}. Sometimes, the polarization of the injected current is reversed at frequencies $\lesssim 100$~Hz to reduce the effect of electrolytic polarization (build up of anions and cations around the electrodes) and telluric currents (currents in the ground mainly produced by the presence and fluctuation of the Earth's magnetosphere) \citep{ieee2011,reynolds2011,clement2020}.}. Despite not expecting a good agreement, such an agreement would be mathematically possible when accounting for the spatial dependence of the results. Our result represents a weighted average of the soil conductivity over space, where the weights corresponds to the sensitivity of the antenna at different coordinates in the soil. Reporting the pattern for the antenna sensitivity in the soil falls outside the scope of this paper. However, using a reasonable set of weights for the conductivities of the four layers of the ERT model it would be possible to obtain a weighted average close to our value. A realistic situation consists of the antenna having higher sensitivity toward the top of the soil. Therefore, assigning as an example weights of $96\%$ to the top layer, $4\%$ to the second layer, and $0\%$ to the other layers, produces a weighted average of $0.0076$~Sm$^{-1}$, consistent with our result $0.0070\pm0.0014$~Sm$^{-1}$.

\begin{table}
\caption{Four-layer soil conductivity model obtained by \citet{alegria2013} for a site $650$~m east of Site A using ERT.}
\label{table_ERT}
\centering
\begin{tabular}{lcc}
\hline
\hline
\noalign{\smallskip}
Layer & Thickness [m] & $\hat{\sigma}$~[Sm$^{-1}$] \\
\noalign{\smallskip}
\hline
\noalign{\smallskip}
1 (top)       & $0.72$ & $0.00239$ \\
2             & $5.4$  & $0.133$  \\
3             & $90$   & $1.19$  \\
4 (bottom)    & infinite  & $0.00011$  \\
\noalign{\smallskip}
\hline
\end{tabular}
\end{table}

\section{Discussion and summary}
Our results show that the impedance of an antenna used to conduct ground-based global $21$~cm observations can be used for the electrical characterization of the soil in the same frequency range. We demonstrate this technique using the antenna of the MIST experiment. The electromagnetic performance of MIST is very sensitive to the soil electrical properties due to MIST's strategy of observing the sky without a metal ground plane under the antenna. Therefore, an accurate characterization of the soil at the observation sites is critical for the experiment's success. The technique demonstrated here is particularly attractive for MIST since the measurement of the antenna impedance is already included in the calibration program of the experiment \citep{monsalve2024a}. 

The antenna impedance technique was used to estimate the soil electrical conductivity and relative permittivity at three test sites in the Greater Concepci\'on area, Chile. The parameter estimates were determined by fitting impedance measurements with models from electromagnetic simulations of the antenna and soil. The soil was simulated as homogeneous. The measurements and simulations were conducted in the range $25$--$125$~MHz. For the three sites, the electrical conductivity was found between $0.007$ and $0.049$~Sm$^{-1}$, and the relative permittivity between $1.6$ and $12.7$. The percent precision of the estimates at $68\%$ probability is, with one exception, better (lower) than $33\%$. The best-fit simulated impedances have a better than $10\%$ agreement with the data relative to the peak absolute values of the resistance and reactance across our frequency range. Interpreting the fit residuals will require knowledge about the measurement uncertainties in future implementations of the technique.

This technique has been used in the past to characterize the environment around an antenna \citep[e.g.,][]{wakita2000,lenlereriksen2004,berthelier2005,legall2006} but, to our knowledge, we are the first to use it in the context of $21$~cm cosmology. Measurements of soil electrical parameters reported so far by $21$~cm experiments have been done using other techniques \citep{sutinjo2015,spinelli2022}. To demonstrate the impedance technique, in this initial effort we have focused on computational aspects, in particular the electromagnetic simulations and parameter estimation. The total number of simulations performed to estimate the parameters was $\leq 602$ for each of the three test sites. Although the simulations are computationally expensive, not having to include a large metal ground plane makes the MIST simulations simpler and faster, which is a critical factor to consider when hundreds or thousands of simulations are required to characterize the soil. 

The soil parameter accuracy required to detect the global $21$~cm signal depends, among other factors, on the properties of the soil itself\footnote{For instance, the accuracy requirement on $\sigma$ is in general a function of $\sigma$ and $\epsilon_r$.}, the characteristics of the instrument, and the data analysis strategy used \citep{spinelli2022,monsalve2024b,pattison2025}. It is therefore difficult to determine accuracy requirements that are applicable beyond a specific experimental scenario and analysis approach. In \citet{monsalve2024b} we estimated first-order soil parameter accuracy requirements for MIST when using a simple cosmological analysis approach and assuming homogeneous and two-layer models for the soil. We found that for homogeneous soils, an accuracy of $10\%$ in $\sigma$ and $\epsilon_r$ was sufficient to obtain reasonably accurate estimates for the global $21$~cm signal parameters (first column of Figure~$1$ in \citet{monsalve2024b}). Two-layer soils generally increased the accuracy requirements to percent level (second through fifth columns of the same figure).

The parameter uncertainties reported in this paper ---in most cases better than $33\%$ at $68\%$ probability--- are interesting considering that \citet{monsalve2024b} hints at the possibility that errors $>10\%$ might be tolerable for homogeneous soils. Nonetheless, acknowledging the limitations of homogeneous soil models and anticipating additional challenges in the analysis of real observations, we aim to significantly increase the accuracy and precision of our soil parameter estimates. To characterize the soil at the sky observation sites, we will use the impedance technique with the full MIST instrument instead of only the antenna panels connected with exposed cables and wires as in this paper. Using the technique with the full instrument will lead to higher accuracy in the measurements and electromagnetic simulations, and consequently higher accuracy in the soil parameters.

The impedance measurements done with the full instrument are expected to be more accurate for three main reasons. First, the instrument carries out the measurements autonomously, thus eliminating the impact of the human presence and operations. Second, the impedance is measured with a better VNA and calibrated with accurately modeled OSL standards, all of which are contained in the receiver box located under the antenna \citep{monsalve2024a}. Third, the instrument does not use external cables or the adapter required for this paper, which introduce time instabilities that cannot be tracked and corrected. The electromagnetic simulations of the full instrument are also expected to be more accurate. The cable and adapter used for this paper have a complicated geometry, which introduces inaccuracies in the simulations. These inaccuracies will be avoided in the simulations of the instrument. Further, the simulation geometry will be designed using accurate measurements of the instrument's physical dimensions and orientation done in the field, instead of relying on nominal values and laboratory measurements as in this paper. 

In addition to increasing the accuracy in the measurements and simulations, quantifying their uncertainties represents a priority for future implementations of this technique. Robust uncertainty estimates for the measurements and simulations will enable us to: (1) use rigorous statistical methods to select between competing models for the soil as we attempt to make the model more realistic, and (2) determine reliable uncertainties for the soil parameters. 

The soil characterization technique presented in this paper is directly relevant for global $21$~cm experiments planning to observe the sky from the surface of the lunar farside \citep[e.g.,][]{bale2023}. The electrical characteristics of the lunar regolith vary significantly with location \citep{feng2020,siegler2020}. Therefore, to reach the calibration accuracy required for $21$~cm cosmology, the conductivity and relative permittivity of the regolith will have to be determined at each observation site. Measurements of the antenna impedance primarily conducted for radiometer calibration could also be used for in-situ regolith characterization.

\begin{acknowledgements}
We thank Roberto Vives and Benjam\'in Fern\'andez for their help with the antenna construction and measurements. We also thank Matheus Pess\^oa for his work during early tests of the antenna impedance technique. We acknowledge support from ANID Fondo 2018 QUIMAL/180003, Fondo 2020 ALMA/ASTRO20-0075, Fondo 2021 QUIMAL/ASTRO21-0053, and Fondo 2022 ALMA/31220012. We acknowledge support from ANID FONDECYT Iniciaci\'on 11221231 and ANID FONDECYT Regular 1251819. We acknowledge support from Universidad Cat\'olica de la Sant\'isima Concepci\'on Fondo UCSC BIP-106. We acknowledge the support of the Natural Sciences and Engineering Research Council of Canada (NSERC), RGPIN-2019-04506, RGPNS 534549-19. We acknowledge the support of the Canadian Space Agency (CSA) [21FAMCGB15]. This research was undertaken, in part, thanks to funding from the Canada 150 Research Chairs Program. This research was enabled in part by support provided by SciNet and the Digital Research Alliance of Canada. 
\end{acknowledgements}

\software{SciPy \citep{virtanen2020} }

\appendix

\section{Geometry of the Feko simulations}
\label{section_appendix_feko}

\subsection{Simple geometry}
\label{section_appendix_simple_geometry}
We initially conducted a set of Feko simulations with a simple geometry to evaluate how well the simulated impedance could fit the measurements without incorporating into the geometry small features that increase the computation time. The simple geometry only consists of the antenna panels, an idealized excitation port, and the soil. The antenna panels are simulated as aluminum boxes of nominal dimensions: $1.2$~m in length, $60$~cm in width, $3$~mm in thickness, and $2$~cm in horizontal separation. The length and width of the panels are aligned parallel to the $xy$-plane at a height $z=50$~cm. The long dimension of the panels is aligned with the $x$-axis. The gap between the panels is centered at $(x,y,z)=(0,0,50\;\mathrm{cm})$. The panels are connected by an ideal $2$~cm-long wire centered at $(x,y,z)=(0,0,50\;\mathrm{cm})$ and aligned with the $x$-axis. A voltage port representing the electrical excitation of the antenna is placed in the middle of the wire. The simulated soil is homogeneous, and its surface is perfectly flat and aligned with the $xy$-plane at $z=0$.

\begin{figure*}
\centering
\includegraphics[width=\hsize]{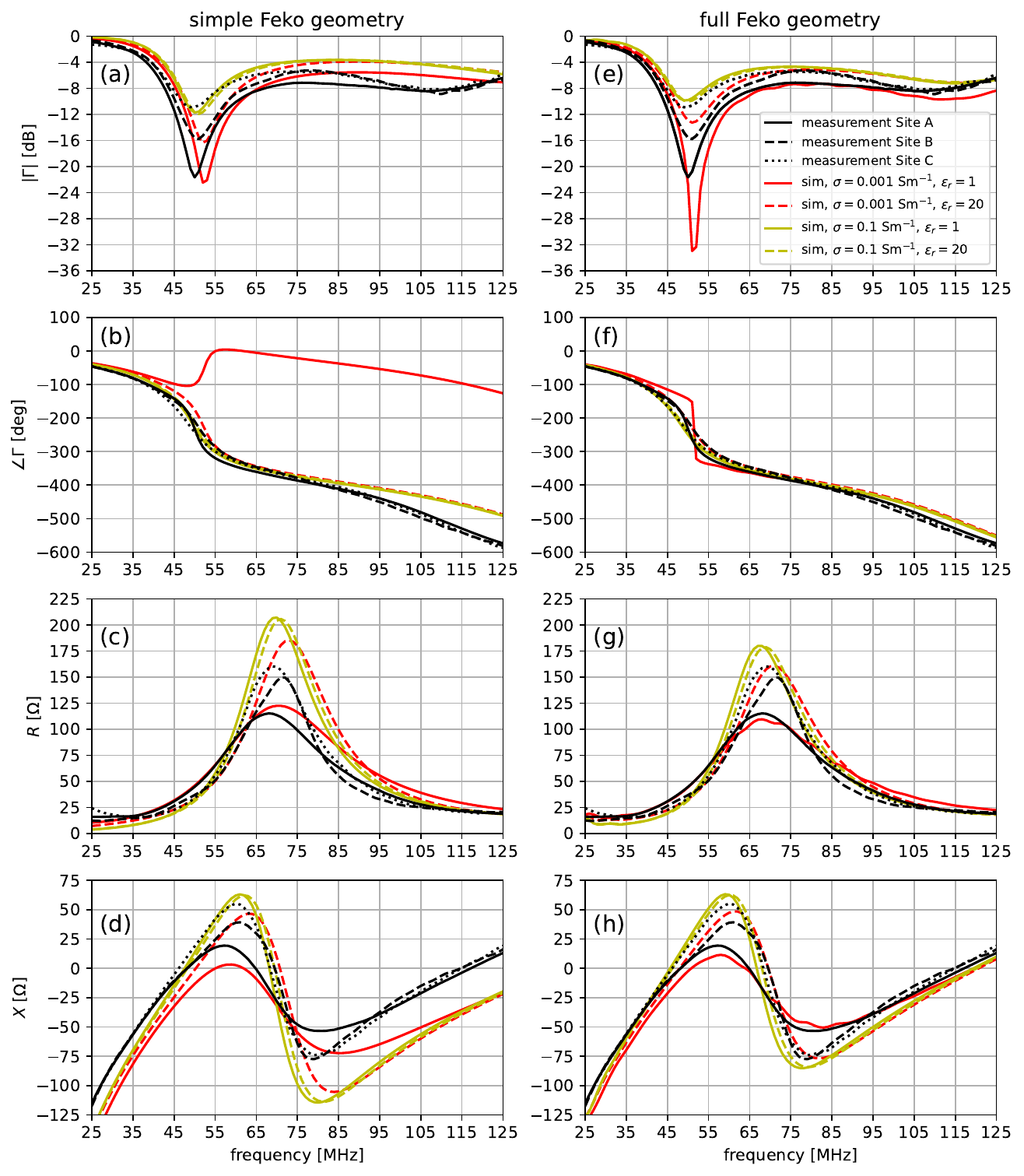}
\caption{Comparison between the impedance measurements and Feko simulations done with (left) a simple geometry and (right) a full geometry of the MIST antenna and measurement setup. The soil parameter combinations used in these simulations span a wide realistic space. The simple simulation geometry ---which only includes the antenna panels, an idealized excitation port, and the soil--- leads to large systematic differences with the data. The full simulation geometry ---which also includes the wires and SMA connector in the gap between the panels, as well as the $20$~m cable--- improves the overall agreement.}
\label{figure_feko_simple_full}
\end{figure*}

Panels (a)--(d) of Figure~\ref{figure_feko_simple_full} show four simulations conducted with the simple geometry, as well as the measurements at the three test sites for reference. In the simulations, the soil was assigned $(\sigma,{\epsilon}_r)=(0.001\;\mathrm{Sm}^{-1},1)$, $(0.001\;\mathrm{Sm}^{-1},20)$, $(0.1\;\mathrm{Sm}^{-1},1)$, and $(0.1\;\mathrm{Sm}^{-1},20)$. These combinations span a wide realistic parameter space. The panels in the figure show that across some frequency ranges the simulations with the simple geometry differ from the measurements by amounts that cannot be attributed to the soil parameters. The clearest example is in $X$, where above $85$~MHz the simulations and measurements converge to different values, which at $125$~MHz are separated by $\sim$$35$~$\Omega$. Systematic differences between simulations and measurements can also be seen in: (1) $|\Gamma|$ at $\sim$$85$--$120$~MHz; (2) $\angle \Gamma$ above $\sim$$85$~MHz; (3) $R$ at $\sim$$80$--$105$~MHz; and (d) $X$ at $\sim$$25$--$45$~MHz.

The systematic differences observed between the simulations with the simple geometry and the data motivated us to make the geometry more realistic. The improved geometry, which we here call `full', is described in Appendix~\ref{section_appendix_full_geometry}. Panels (e)--(h) of Figure~\ref{figure_feko_simple_full} show simulations with the full geometry. In these simulations, the soil parameters take the same values as in the simulations with the simple geometry shown in panels (a)--(d). The simulations with the full geometry agree better with the measurements, and some remaining systematic differences are significantly smaller than when using the simple geometry. These results made us choose the full geometry for the simulations in this paper.

\subsection{Full geometry}
\label{section_appendix_full_geometry}

In the full simulation geometry, the antenna panels and soil have the same characteristics as in the simple geometry. However, we remove the idealized excitation port from the gap between the panels. We then incorporate the wires and SMA connector of the adapter between the panels, as well as the $20$~m coaxial cable.

The wires of the adapter are simulated as copper cylinders with a diameter of $0.64$~mm. The wires make contact with the panels while aligned with the $x$-axis at the center of the gap. As they leave the panels, the wires bend in the $+y$~direction, where they extend for $6.3$~cm before reaching the SMA connector. The connector is simulated as a gold cuboid of $1.6$~cm in length and $6$~mm in width and thickness. The electrical excitation of the antenna is applied by a voltage port located at the junction between one of the wires and the SMA connector. The location of the port mimics the impedance calibration point for the real antenna at the SMA connector.

The cable is simulated as a $2$~mm  diameter copper cylinder leaving the antenna in the $+y$ direction. This cylinder begins at the output of the SMA connector and runs along the middle of the gap between the panels up to the edge of the panels. At the edge of the panels, the cylinder bends downward $18.3^{\circ}$ and continues in diagonal until it reaches a lowest height of $5$~mm above the soil. This point occurs at a distance of $1.5$~m from the edge of the panels in the $+y$~direction. The cylinder then starts running parallel to the soil in the $+y$~direction at a height of $5$~mm until it reaches a total length of $20$~m.

In the simulations we use the following conductivities \citep{pozar2005}: for aluminum, $3.816\times 10^{7}$~Sm$^{-1}$; for copper, $5.813\times 10^{7}$~Sm$^{-1}$; and for gold, $4.098\times 10^{7}$~Sm$^{-1}$. The support structure of the antenna, made out of fiberglass and plastic, does not affect the antenna impedance and is not included in the simulations. Other features not included are the mounting holes of the antenna panels. These holes would significantly increase the complexity of the geometry and computation time but according to our tests do not have a noticeable effect on the impedance.

\section{Estimating the soil parameters and their uncertainties}
\label{section_appendix_fitting_strategy}

To determine the best-fit soil parameter values we use the simplicial homology global optimization (SHGO) algorithm \citep{endres2018, virtanen2020}. SHGO enables derivative-free, constrained and unconstrained optimization. SHGO minimizes the cost function using global and local searches. For the global search, SHGO incorporates the simplicial \citep{hatcher2002}, Sobol \citep{sobol1967, joe2008}, and Halton \citep{halton1960} sampling methods, and can also accept externally provided sampling points. In this work we use the Sobol and Halton low-discrepancy sequences for the global search. These sequences are deterministic sets of sampling points with mathematical properties comparable to, and in some cases better than, the points from the popular Latin hypercube sampling \citep{mckay1979,renardy2021}. For the local search, we use sequential least squares quadratic programming \citep{kraft1988, nocedal2006}, which is the default method in SHGO.

\begin{figure*}
\centering
\includegraphics[width=\hsize]{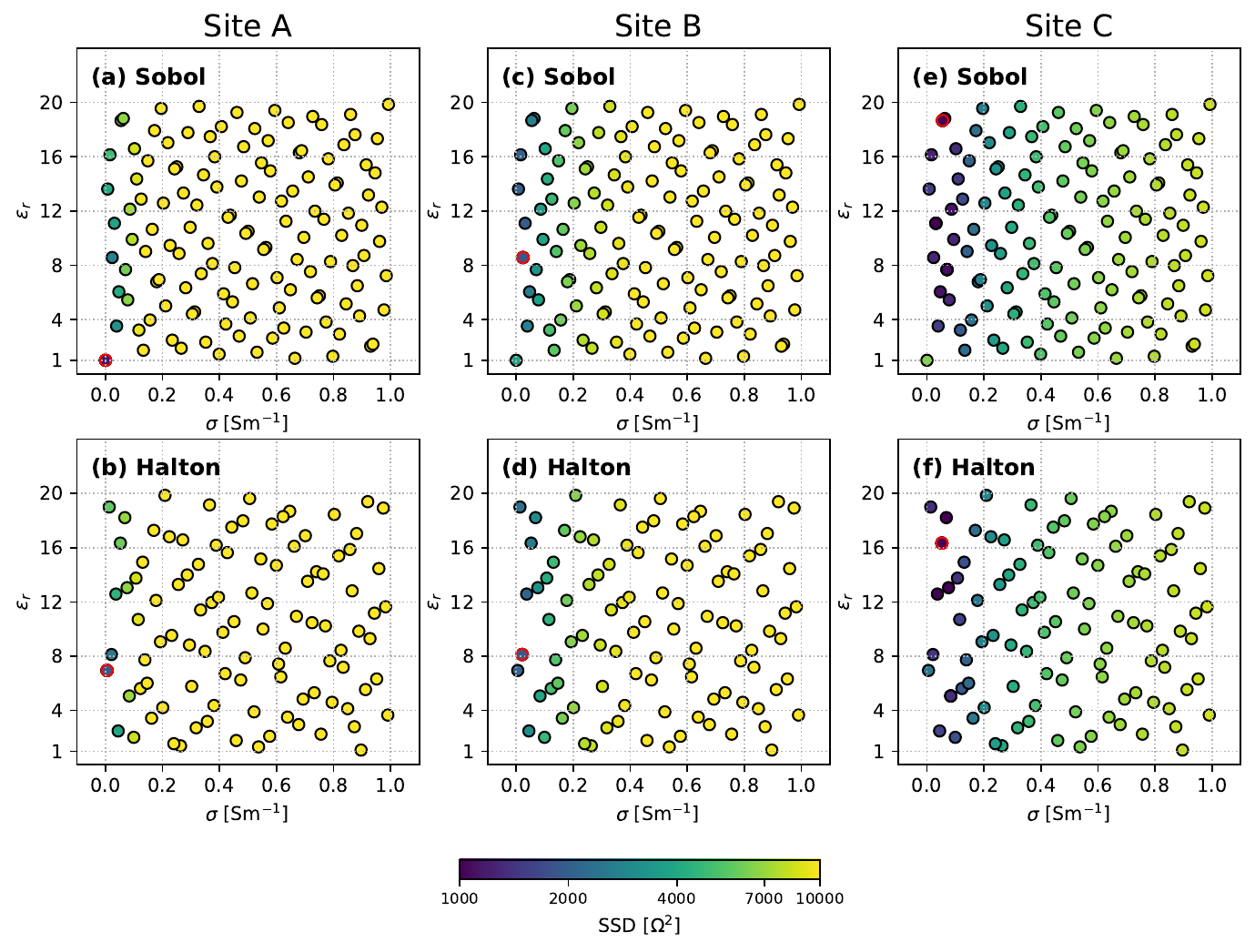}
\caption{Samples produced in the initial fitting step. The colors of the samples represent the SSD. The three columns correspond to the three test sites. The top (bottom) row corresponds to the Sobol (Halton) samples. In each panel, the sample with the lowest SSD, denoted by SSD$^{\star}$, is highlighted using a red circle. In all cases, SSD$^{\star}$ occurs for $\sigma<0.06$~Sm$^{-1}$.}
\label{figure_samples_start}
\end{figure*}

\begin{figure*}
\centering
\includegraphics[width=\hsize]{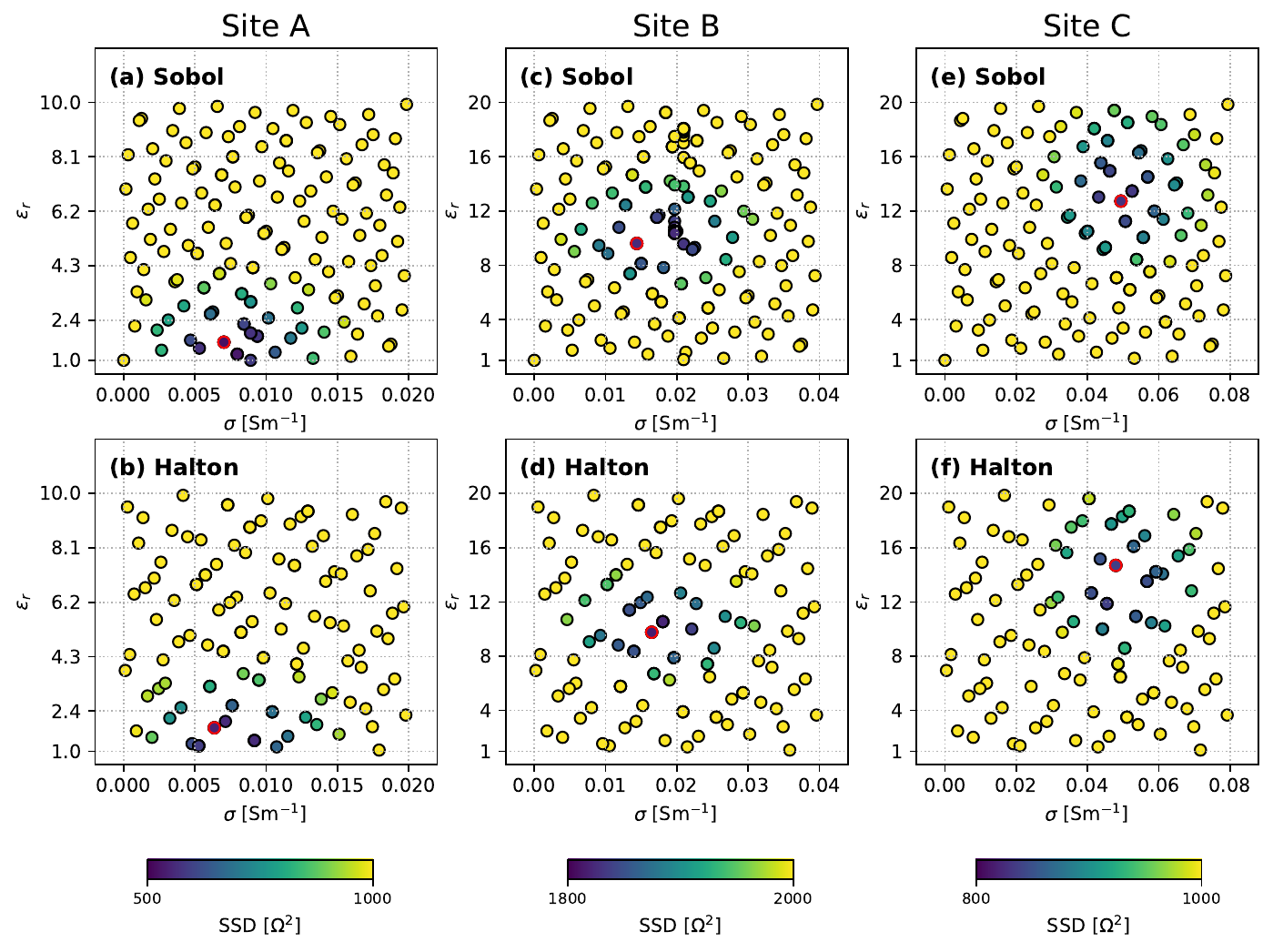}
\caption{Same as Fig.~\ref{figure_samples_start} but for the final fitting step. This step enables the identification of the best-fit values with higher accuracy than the initial step. The color scales represent different SSD ranges for each of the three sites.}
\label{figure_samples_end}
\end{figure*}

During the parameter fits we use the default settings that define the stopping criterion in SHGO. With this choice it is necessary to fit the parameters in two steps in order to explore a wide parameter space and converge to accurate best-fit values. In the initial step we run SHGO within the wide space defined by the following parameter ranges: $0$--$1$~Sm$^{-1}$ for $\sigma$ and $1$--$20$ for $\epsilon_r$. The best-fit values found in this step are not very accurate but indicate the regions where the true best-fit values lie. In the final fitting step we run SHGO again but within smaller parameter spaces, which we select based on the best-fit values found in the initial step. To quantify the accuracy of the best-fit values we carry out the two-step fitting process twice, using separately the Sobol and Halton sampling points for the global exploration. Although conducting the fits twice roughly doubles the number of Feko evaluations, the total number is expected to be less than 1000 per site, which is lower than typical numbers required for MCMC or nested sampling.

The fit results are shown in Figures~\ref{figure_samples_start} and \ref{figure_samples_end}, and summarized in Table~\ref{table_results}. The table lists: (1) the best-fit soil parameters, $\hat{\sigma}$ and $\hat{\epsilon}_r$; (2) the sum of square deviations (SSD) at the best-fit values, SSD$^{\star}$; (3) the root mean square deviation (RMSD) between the best-fit impedance models and the data, $\mathrm{RMSD}^{\star}=\sqrt{\mathrm{SSD}^{\star}/N}$, where $N=22$; and (4) the number of evaluations, $N_{\mathrm{ev}}$. These results are presented for the initial and final fitting steps, and for the Sobol and Halton samples.

Figure~\ref{figure_samples_start} shows the samples and SSDs produced in the initial fitting step. The SSDs go from under 2000~$\Omega^2$ to over 7000~$\Omega^2$. For the three test sites, the SSD decreases toward low $\sigma$, while across $\epsilon_r$ it remains more stable. The samples that produce the SSD$^{\star}$ are indicated using red circles. For Site~A, as shown in Table~\ref{table_results}, the best-fit values found in the initial step are $(\hat{\sigma},\hat{\epsilon}_r)=(10^{-5}\;\mathrm{Sm}^{-1},1)$ for Sobol and $(0.0054\;\mathrm{Sm}^{-1},6.95)$ for Halton. These values are used as reference to determine adequate parameter ranges for the final fitting step. The ranges selected for the final step are $0$--$0.02$~Sm$^{-1}$ for $\sigma$ and $1$--$10$ for $\epsilon_r$, which contain the initial best-fit values. For Sites~B and C we follow a similar approach but decide to only restrict the $\sigma$ range. For these sites, the best-fit values found for $\epsilon_r$ in the initial step are higher than for Site~A and the variation of the SSD across $\epsilon_r$ is low. We thus prefer to maintain $1$--$20$ as the range for $\epsilon_r$ in the final step. The $\sigma$ range is restricted to $0$--$0.04$~Sm$^{-1}$ for Site~B and $0$--$0.08$~Sm$^{-1}$ for Site~C.

\begin{table*}
\caption{Results of the parameter estimation with SHGO.}
\label{table_results}
\centering
\begin{tabular}{clccccccccccc}
\hline
\hline
\noalign{\smallskip}
&  & \multicolumn{5}{c}{Initial fitting step} & \multicolumn{5}{c}{Final fitting step} \\
\cmidrule(lr){3-7}
\cmidrule(lr){8-12}
Site & Sampling  & $\hat{\sigma}$~[Sm$^{-1}$] & $\hat{\epsilon}_r$ & SSD$^{\star}$~[$\Omega^2$] & RMSD$^{\star}$~[$\Omega$] & $N_{\mathrm{ev}}$ & $\hat{\sigma}$~[Sm$^{-1}$] & $\hat{\epsilon}_r$ & SSD$^{\star}$~[$\Omega^2$] & RMSD$^{\star}$~[$\Omega$] & $N_{\mathrm{ev}}$ \\
\noalign{\smallskip}
\hline
\noalign{\smallskip}
A     & Sobol   & $10^{-5}$ &     $1$ & $1132.0$ & $7.2$ & $131$ & $0.0070$ &  $1.63$ &  $513.3$ & $4.8$ & $183$ \\
A     & Halton  & $0.0054$  &  $6.95$ & $1880.6$ & $9.2$ & $103$ & $0.0064$ &  $1.82$ &  $531.8$ & $4.9$ & $151$ \\
B     & Sobol   & $0.0234$  &  $8.57$ & $1869.1$ & $9.2$ & $131$ & $0.0144$ &  $9.61$ & $1822.6$ & $9.1$ & $220$ \\
B     & Halton  & $0.0210$  &  $8.12$ & $1861.1$ & $9.2$ & $103$ & $0.0165$ &  $9.77$ & $1815.5$ & $9.1$ & $148$ \\
C     & Sobol   & $0.0547$  & $18.66$ & $930.0$  & $6.5$ & $137$ & $0.0494$ & $12.73$ &  $825.9$ & $6.1$ & $188$ \\
C     & Halton  & $0.0523$  & $16.33$ & $864.2$  & $6.3$ & $106$ & $0.0479$ & $14.69$ &  $834.1$ & $6.2$ & $148$ \\
\noalign{\smallskip}
\hline
\end{tabular}
\end{table*}

Figure~\ref{figure_samples_end} shows the samples and SSDs produced in the final fitting step. The final parameter spaces selected based on the initial fitting step successfully capture the regions of lowest SSD and enable a denser sampling of these regions. The denser sampling leads to a higher accuracy in the best-fit values, which is reflected in the agreement at the $\sim$$15\%$ level or better between the Sobol and Halton best-fit values. We report as the final best-fit values those that produce the lowest SSD considering both the Sobol and Halton sample sets. For Sites~A, B, and C, the final best-fit values are $(\hat{\sigma},\hat{\epsilon}_r)=(0.0070\;\mathrm{Sm}^{-1},1.63)$, $(0.0165\;\mathrm{Sm}^{-1},9.77)$, and $(0.0494\;\mathrm{Sm}^{-1},12.73)$, respectively. The corresponding SSD$^{\star}$ are $513.3$, $1815.5$, and $825.9$~$\Omega^2$. The total number of sampled points considering the two fitting steps and the Sobol and Halton samples is $\leq602$ for each of the three sites.

\bibliography{bibtex}{}
\bibliographystyle{aasjournal}



\end{document}